\documentstyle[aps,prl,twocolumn]{revtex}
\begin{document}
\input{psfig}
\twocolumn[\hsize\textwidth\columnwidth\hsize\csname 
@twocolumnfalse\endcsname 

\title{Soft Mode Anomalies in the Perovskite Relaxor
  Pb(Mg$_{1/3}$Nb$_{2/3}$)O$_3$}

\author{ P.\ M.\ Gehring$^{(a)}$, S.\ B.\ Vakhrushev$^{(b)}$, G.\ Shirane$^{(c)}$ }

\address{ $^{(a)}$NIST Center for Neutron Research, National Institute
of Standards and Technology, Gaithersburg, Maryland 20899 }

\address{ $^{(b)}$Ioffe Phys.\ -Tech.\ Institute, 194021, St.\ 
  -Petersburg, Russia}

\address{ $^{(c)}$Physics Department, Brookhaven National Laboratory,
Upton, New York 11973 }

\maketitle

\begin{abstract}
  Neutron inelastic scattering measurements of the polar TO phonon
  mode in the cubic relaxor Pb(Mg$_{1/3}$Nb$_{2/3}$)O$_3$, at room
  temperature, reveal anomalous behavior similar to that recently
  observed in Pb(Zn$_{1/3}$Nb$_{2/3}$)$_{0.92}$Ti$_{0.08}$O$_3$ in
  which the optic branch {\it appears} to drop precipitously into the
  acoustic branch at a finite value of the momentum transfer $q =
  0.2$~\AA$^{-1}$, measured from the zone center.  By contrast, a
  recent neutron study indicates that PMN exhibits a normal TO phonon
  dispersion at 800~K.  We speculate this behavior is common to all
  relaxor materials, and is the result of the presence of
  nanometer-scale polarized domains in the crystal that form below a
  temperature $T_d$, which effectively prevent the propagation of long
  wavelength ($q=0$) phonons.
\end{abstract}


]

\section{Introduction}

In the past year two neutron inelastic scattering studies have been
carried out in an attempt to elucidate the nature of the lattice
dynamics in the relaxor-based systems Pb(Mg$_{1/3}$Nb$_{2/3}$)O$_3$
(PMN) and Pb(Zn$_{1/3}$Nb$_{2/3}$)$_{0.92}$Ti$_{0.08}$O$_3$
(PZN-8\%PT) which have the complex perovskite structure $A(B'B'')O_3$
and $A(B'B''B''')O_3$, respectively \cite{Naberezhnov,Gehring}.  Of
particular interest to both studies was the soft phonon mode that is
ubiquitous in ferroelectric and perovskite systems.  The displacive
phase transition in classical ferroelectric systems such as PbTiO$_3$,
which has the simple $ABO_3$ perovskite structure, is driven by the
condensation or softening of a zone-center transverse optic (TO)
phonon, i.\ e.\ a ``soft mode,'' that transforms the system from a
cubic paraelectric phase to a tetragonal ferroelectric phase.  Direct
evidence of this soft mode behavior is easily obtained from neutron
inelastic scattering measurements made at different temperatures above
the Curie temperature $T_c$.  The top panel of Fig.~1 shows, for
example, the phonon dispersion of the lowest-energy TO branch in
PbTiO$_3$ (PT) at 20~K above $T_c$.  Here one can see that the zone
center ($\zeta = 0$) phonon energy has already dropped to a very low
value of 3~meV \cite{Shirane}.  As $T \rightarrow T_c$, the soft mode
energy $\hbar \omega_o \propto (T-T_c)^{1/2} \rightarrow 0$.

By contrast, the so-called ``relaxor'' systems possess a built-in
disorder that stifles the ferroelectric transition.  Instead of a
sharp transition at $T_c$, one observes a diffuse phase transition in
which the dielectric permittivity $\epsilon$ exhibits a broad maximum
as a function of temperature at a temperature $T_{max}$.  The disorder
in relaxors can be either compositional or frustrated in nature.  In
the case of PMN and PZN (Z = Zn), the disorder results from the
$B$-site being occupied by ions of differing valence (either Mg$^{2+}$
or Zn$^{2+}$, and Nb$^{5+}$).  Hence the randomness of the $B$-site
cation breaks the translational symmetry of the crystal.  Yet despite
years of intensive research, the physics of the observed diffuse phase
transition is still not well understood \cite{Westphal,Colla,Blinc}.
Moreover, it is interesting to note that no definitive evidence for a
soft mode has been found in these systems.

%
\vspace{0.15in} 
\noindent 
\parbox[b]{3.4in}{ \psfig{file=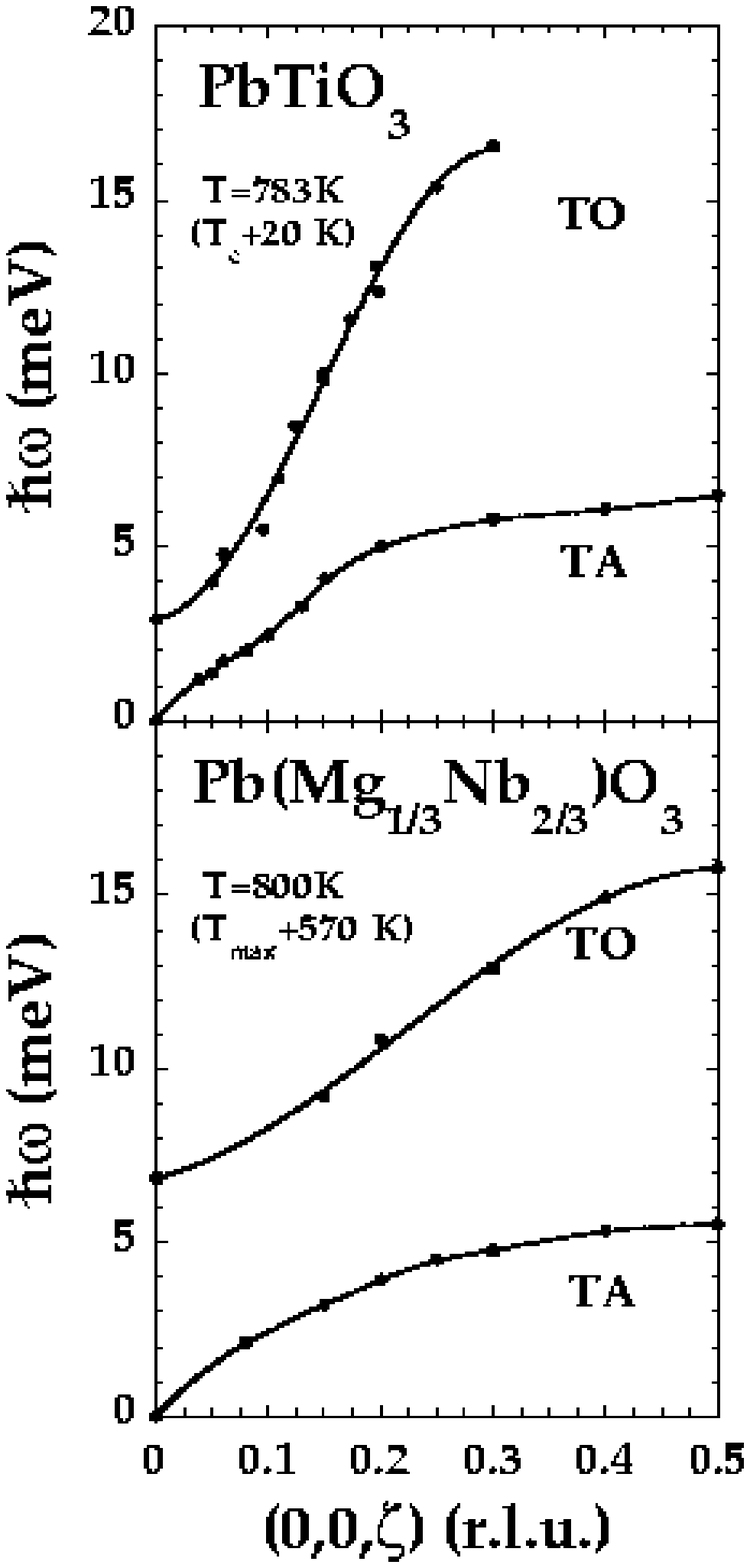,width=2.55in} {Fig.~1. \small
    Top - Dispersions of the lowest-energy TO mode and the TA mode in
    PbTiO$_3$, measured just above $T_c$ (from \cite{Shirane}).
    Bottom - Dispersions of the equivalent modes in PMN measured far
    above $T_{max}$ (from \cite{Naberezhnov}). }}
\vspace{0.05in}
%

In a series of papers published in 1983, Burns and Dacol proposed an
elegant model to describe the disorder intrinsic to relaxor systems
\cite{Burns}. Using measurements of the optic index of refraction on
both ceramic samples of (Pb$_{1-3x/2}$La$_x$)(Zr$_y$Ti$_{1-y}$)O$_3$
(PLZT) as well as microscopically homogeneous single crystals of PMN
and PZN \cite{Burns}, they demonstrated that a randomly-oriented local
polarization $P_d$ develops at a well-defined temperature $T_d$, often
referred to as the Burns temperature, several hundred degrees above
the apparent ferroelectric transition temperature $T_c$.  The spatial
extent of these locally polarized regions in the vicinity of $T_d$ was
conjectured to be $\sim$ several unit cells, and has given rise to the
term ``polar micro-regions,'' or PMR \cite{Tsurumi}. For PMN, the
formation of the PMR occurs at $\sim$ 617~K \cite{Burns}, well above
the temperature $T_{max}$ = 230~K where the dielectric permittivity
reaches a maximum \cite{Naberezhnov}.  Recently, using neutron
inelastic scattering techniques, we have found striking anomalies in
the lowest-energy TO phonon branch (the same branch that goes soft at
at the zone center at $T_c$ in PbTiO$_3$) that we speculate are
directly caused by these same nanometer-sized PMR.

\section{Search for a Soft Mode in PMN}

Our phonon measurements on relaxor systems began with PMN at the NIST
Center for Neutron Research (NCNR) in 1997.  At that time many diffuse
scattering studies of PMN using X-rays and neutrons had already been
published \cite{Vakhrushev,You}. However, there were no published
neutron inelastic scattering measurements on PMN until the 1999 phonon
study by Naberezhnov {\it et al} \cite{Naberezhnov}. The bottom panel
of Fig.~1 shows neutron scattering data taken by Naberezhnov {\it et
  al.} on PMN exactly analogous to that shown in the top panel for
PbTiO$_3$, except that the PMN data were taken at 800~K, a temperature
that is much higher relative to the transition temperature of PMN,
i.~e.\ $\sim 570$~K above $T_{max}$.

The neutron scattering measurements presented here were performed at
the NCNR using both the BT2 and BT9 triple-axis spectrometers.  The
(002) reflections of highly-oriented pyrolytic graphite (HOPG)
crystals were used to both monochromate and analyze the incident and
scattered neutron beams.  An HOPG transmission filter was used to
eliminate higher-order neutron wavelengths.  Inelastic measurements
were made by holding the final neutron energy $E_f$ fixed at 14.7~meV
($\lambda_f = 2.36$~\AA) while varying the incident neutron energy
$E_i$.  Typical horizontal beam collimations used were
60$'$-40$'$-40$'$-80$'$ and 40$'$-48$'$-48$'$-80$'$.  The single
crystal of PMN used in this study measures 0.5~cm$^3$ in volume, and
was the identical crystal used by Naberezhnov {\it et al}.  It was
grown using the Chochralsky technique described elsewhere
\cite{Naberezhnov}. The crystal was mounted onto an aluminum holder and
oriented in air with the either the cubic [$\bar{1}$10] or [001] axis
vertical.

We used two types of scans to collect data.  Constant energy
(constant-$E$) scans were performed by keeping the energy transfer
$\hbar \omega = \Delta E = E_f - E_i$ fixed while varying the momentum
transfer $\vec{Q}$.  Constant-$\vec{Q}$ scans were performed by
holding the momentum transfer $\vec{Q} = \vec{k_f} - \vec{k_i}$ ($k =
2\pi/\lambda$) fixed while varying the energy transfer $\Delta E$.
Using these scans, the dispersions of both the transverse acoustic
(TA) and the lowest-energy transverse optic (TO) phonon modes were
mapped out at room temperature (still in the cubic phase, but well
below the Burns temperature $T_d \sim$ 617~K).

%
\vspace{0.15in} 
\noindent 
\parbox[b]{3.4in}{ \psfig{file=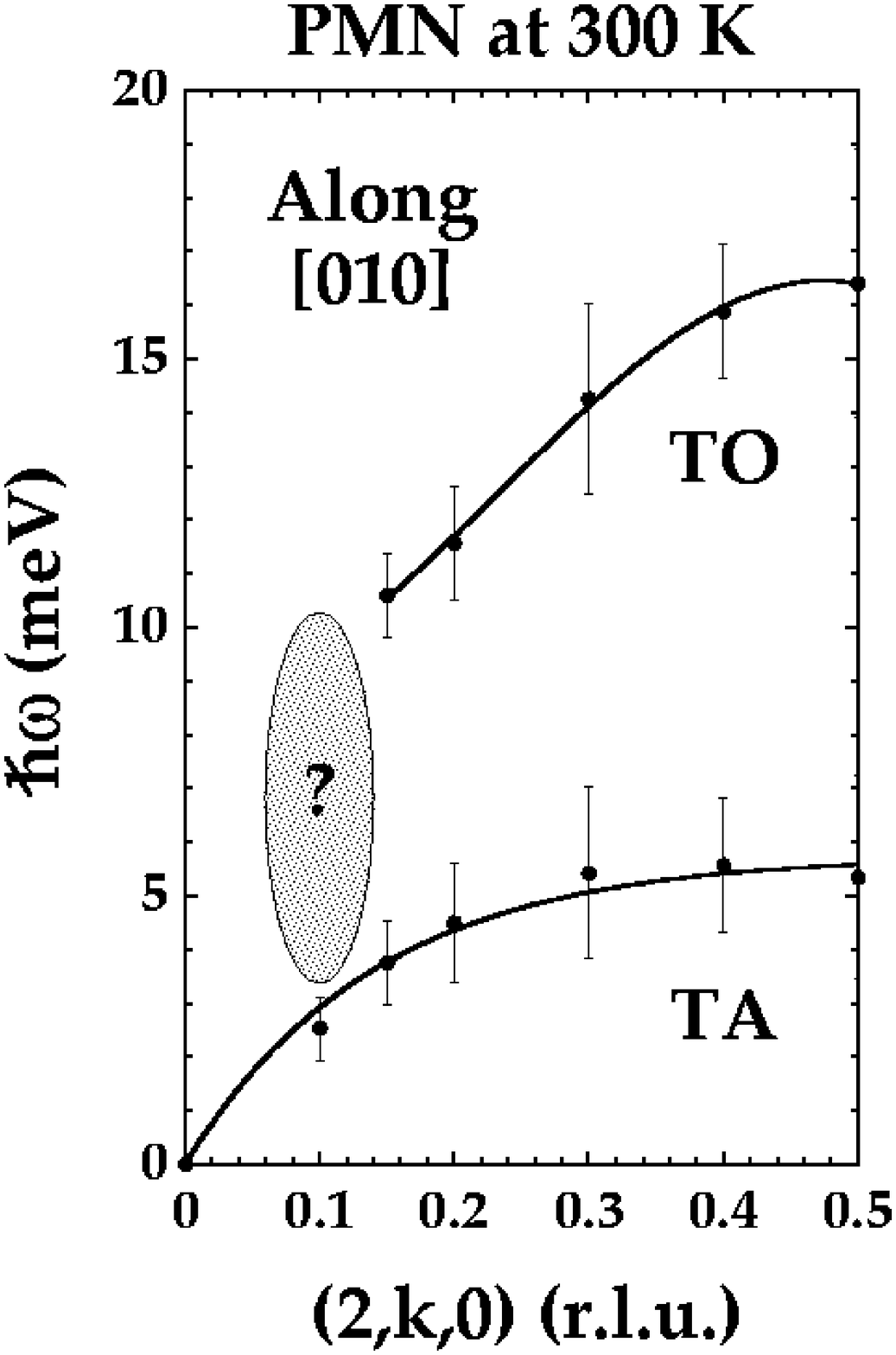,width=2.95in} {Fig.~2. \small
    Solid dots represent positions of peak scattered neutron intensity
    taken from constant-$\vec{Q}$ scans at 300~K along the [010]
    symmetry direction.  Vertical bars represent phonon FWHM
    linewidths in meV.  Solid lines are guides to the eye indicating
    the TA and TO dispersion curves.  Shaded area represents region of
    TO dispersion where constant-$\vec{Q}$ scans showed no
    well-defined peaks. }}
\vspace{0.05in}
%

In Fig.~2 we plot the positions of the peak in the scattered neutron
intensity taken from constant-$\vec{Q}$ scans at 300~K as a function
of $\hbar \omega$ and $|\vec{q}| = k$.  Here $\vec{q} = \vec{Q} -
\vec{G}$ is the momentum transfer measured relative to the $\vec{G} =
(2,0,0)$ Bragg reflection along the [010] symmetry direction.  Limited
data were also taken near $(3,0,0)$.  The lengths of the vertical bars
represent the measured phonon peak FWHM linewidths (full width at half
maximum) in $\hbar \omega$ (meV), and were derived from Gaussian
least-squares fits to the constant-$\vec{Q}$ scans.  The lowest-energy
data points trace out the TA phonon branch along [010], and solid
lines have been drawn through these points as a guide to the eye.  We
see that the TA dispersion curve is identical to that shown for PMN at
800~K in the bottom panel of Fig.~1.

%
\vspace{0.15in} 
\noindent 
\parbox[b]{3.4in}{ \psfig{file=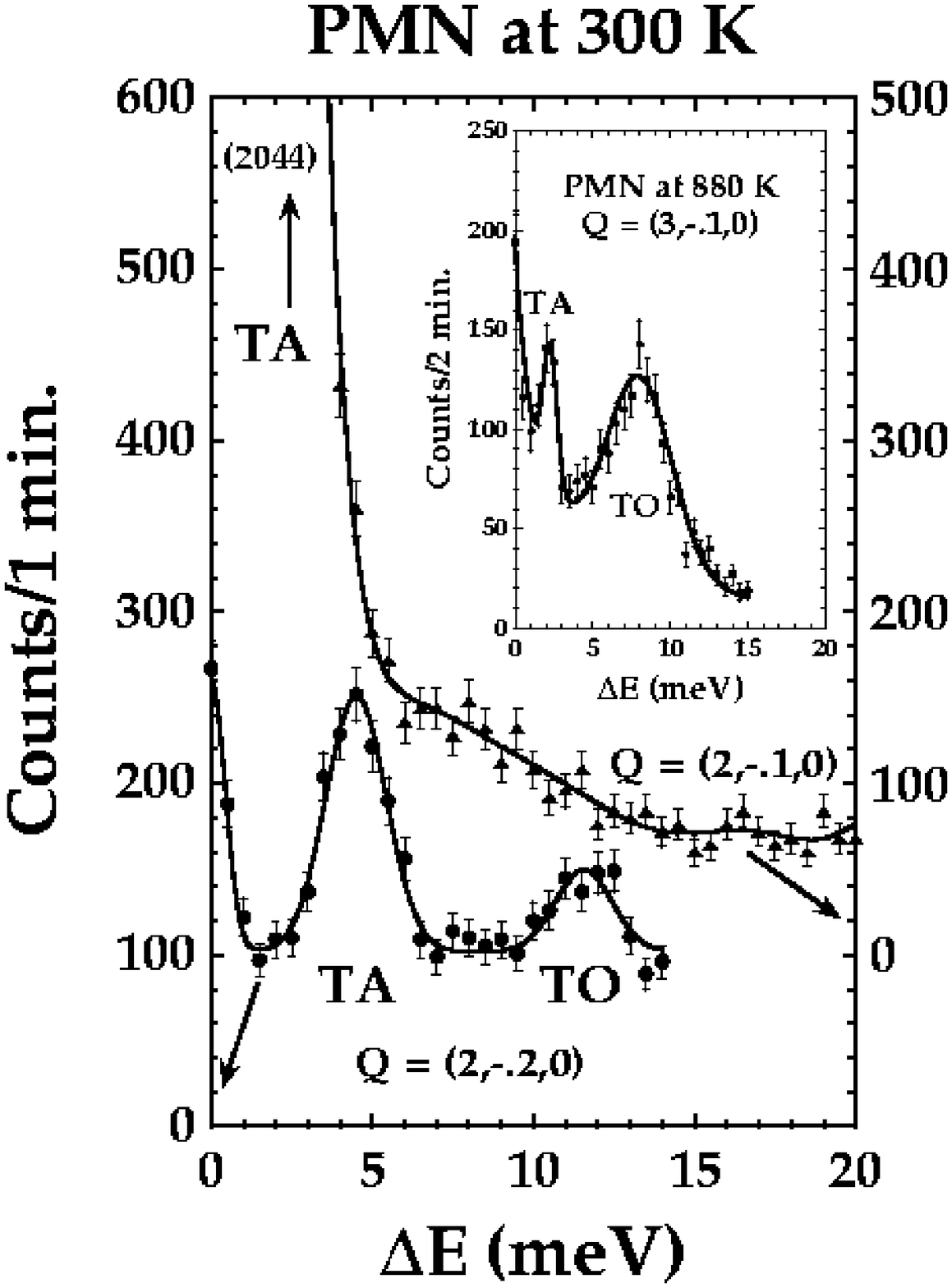,width=3.35in} {Fig.~3. \small
    Data from constant-$\vec{Q}$ scans taken near (2,0,0) at 300~K.
    Lines are guides to the eye.  The scan at $q = -0.2$ r.l.u. shows
    well-defined TA and TO modes.  But at $q = -0.1$ r.l.u., only the
    TA peak is well-defined.  The TO mode is strongly overdamped.  The
    inset, however, shows data taken on the same crystal at the same
    $q$ at 880~K in which the TO mode is clearly well-defined (from
    \cite{Sergey}). }}
\vspace{0.05in}
%

It is also clear from the dispersion diagram presented in Fig.~2 that
our room temperature data show the same TO1 modes at high $q$ as those
reported at 800K by Naberezhnov {\it et al}.  However the scattering
intensities for this mode for small $q \le 0.15$~r.l.u.\ were scarcely
above background at $(2,q,0)$ as well as at $(3,q,0)$.  This is
evident in Fig.~3 where two constant-$\vec{Q}$ scans, taken near
(2,0,0) at 300~K, are shown.  For $q$ = -0.2~r.l.u.  (1~r.l.u.\ =
2$\pi$/a = 1.553~\AA$^{-1}$), we observe two well-defined peaks
corresponding to scattering from the TA and TO modes.  But for $q$ =
-0.1~r.l.u.\ only the TA mode is well-defined.  The TO mode scattering
is weak and broadly distributed in $q$.  By contrast, the inset of
Fig.~3 shows a very prominent peak in the scattering from the TO mode
at the same $q$ (-0.1~r.l.u.) taken on the same crystal (data from
\cite{Sergey}.  The only difference was that these data were taken at
much higher temperature, i.~e.\ 880~K.  These data remained a puzzle
as we could not locate where the scattering intensity had gone, and we
were forced to abandon our search for the soft mode for the time
being.

\section{The Morphotropic Phase Boundary and PZN-8\%PT}

Our phonon studies of relaxor single crystals were subsequently
resumed two years later from a very different perspective.  The nearly
vertical morphotropic phase boundary (MPB) in
Pb(Zr$_{1-x}$Ti$_x$)O$_3$ (PZT), which separates the rhombohedral and
tetragonal regions of the PZT phase diagram near a Ti concentration of
50\%, had recently been reinvestigated by Noheda {\it et al}
\cite{Noheda,Guo}.  There they found a previously unknown monoclinic
phase above 300~K that separates the tetragonal and rhombohedral
phases.  This was an exceedingly important discovery, and was
extensively discussed in this conference \cite{Noheda_aspen}, because
the monoclinic phase forms a natural bridge between the tetragonal and
rhombohedral phases, and sheds new light on possible explanations for
the enhanced piezoelectric properties observed in PZT ceramics for
compositions that lie close to the MPB.

Motivated by this result, Gehring, Park and Shirane realized that a
very similar MPB boundary exists in $(1-x)$PZN-$x$PT around $x=0.08$
\cite{Gehring}.  These are solid solutions which exhibit even greater
piezoelectric properties than are obtained in PZT ceramics.  Moreover,
unlike the case of PZT, they can be grown into large high quality
single crystals, ideal for neutron inelastic scattering studies.  The
soft phonons in PbTiO$_3$ (PT) had already been thoroughly
characterized by Shirane {\it et al} \cite{Shirane}.  Hence their idea
was to study the phonons in $(1-x)$PZN-$x$PT at higher $x$, say 20\%
PT, and then trace the evolution of the transverse optic modes to 8\%
PT and PZN as a function of $x$.  In this way it was hoped that the
scattering associated with the missing optic branch at small $q$ for
PMN could be located.

The neutron inelastic measurements on PZN-8\%PT were performed at
500~K.  By employing a combination of both constant-$\vec{Q}$ and
constant-$E$ scans, an anomalous enhancement of the scattering cross
section was discovered between 0.10~r.l.u. $< |\vec{q}| <$
0.15~r.l.u.\ \cite{Gehring}. This enhancement was located at a fixed
$q$ relative to the zone center, and was energy independent over a
large range of energy transfer extending from 4~meV to 9~meV.  When
plotted in the form of a ``dispersion'' diagram, the TO branch
appeared to drop precipitously into the acoustic branch.  This
resulting TO branch was referred to as a ``waterfall'' for this
reason, and is shown in the inset to Fig.~4.

It was conjectured that these waterfalls are caused by the polar
micro-regions first demonstrated by Burns and Dacol \cite{Burns}. The
existence of such polarized regions, which are of finite spatial
extent, should effectively inhibit the propagation of the
ferroelectric TO mode.  Moreover, the size of these regions can be
estimated as $2\pi/q$, which at 500~K corresponds to about 31~\AA, or
roughly 7 to 8 unit cells. This value is consistent with that put
forth in the picture of Burns and Dacol.

%
\vspace{0.15in} 
\noindent 
\parbox[b]{3.4in}{ \psfig{file=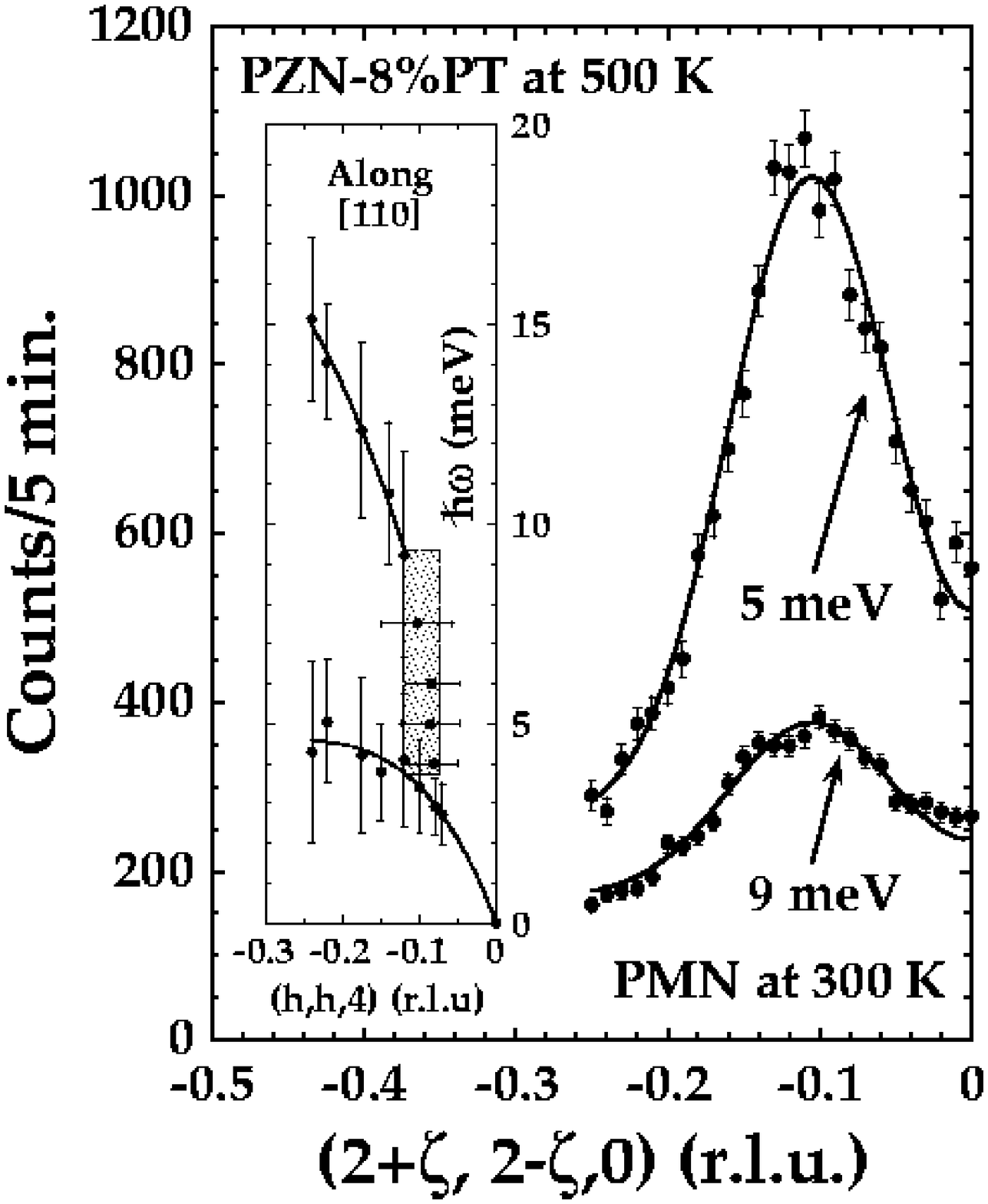,width=3.35in} {Fig.~4. \small
    Two constant-E scans measured along [110] at 5~meV and 9~meV at
    300~K.  These scans demonstrate the presence of the same anomalous
    scattering that was observed in PZN-8\%PT (shown in the inset,
    from \cite{Gehring}). }}
\vspace{0.05in}
%

At this point it was natural to ask the question whether or not the
same anomalous ``waterfall'' was present in PMN as this would provide
a natural explanation for the missing low-$q$ portion of the optic
branch in PMN.  So we went back to our old data of PMN taken at room
temperature, and we found two constant-$E$ scans at 5 and 9~meV which
we had taken along the [110] direction from near (2,2,0).  These data,
shown in the right hand portion of Fig.~4, clearly indicate the
presence of an anomalous enhancement of the scattering intensity at a
fixed $q$, similar to that observed in PZN-8\%PT.

\section{Discussion and Interpretation}

Naberezhnov {\it et al.} identified the normal-looking optic phonon
branch at 800~K shown in Fig.~1 as a hard TO1 mode, and not the
ferroelectric soft mode, because the $\vec{Q}$-dependence of the
associated dynamic structure factor was inconsistent with that
expected for ferroelectric fluctuations, i.~e.\ nearly no critical
scattering was observed near the (2,2,0) Bragg peak in the vicinity of
$T_{max}$ \cite{Naberezhnov}.  On the other hand, the absence of
critical fluctuations at (2,2,0) may simply mean that the eigenvectors
for PMN are different from those of PbTiO$_3$.  The lowest polar optic
mode is still clearly present and well-defined at 880~K.  At lower
temperatures, using the same PMN single crystal, we observe an
overdamped phonon scattering cross section in addition to this new
anomalous scattering at small $q$ below $T_d$, i.~e.\ the waterfall.
So the proper question to ask is whether or not this TO1 branch is the
lowest-energy polar optic mode in PMN.

Before this point can be settled uniquely, more neutron measurements
will be needed at temperatures both above and below $T_d$ to show
precisely how the anomalous scattering changes with temperature, that
is, whether or not the waterfall evolves into the TO1 branch measured
by Naberezhnov {\it et al}.\ at 880~K.  For this purpose, constant-$E$
scans will be of particular importance since, as we learned in 1997,
the waterfall is not readily visible without them.

At present, our picture of PMN is that at high temperatures $T > T_d$
the system behaves like all other simple perovskites.  When the PMR
are formed below the Burns temperature, the crystal behaves like a
two-phase mixture from a lattice dynamical point of view.  The PMR
exhibit the anomalous waterfall as found in PZN-8\%PT, whereas the
non-PMR regions show a gradual change from the regular TO branch to
one which is overdamped.  These are shown very nicely in the
constant-$\vec{Q}$ data of Vakhrushev {\it et al.} at $(3,q,0)$
between 880~K and 450~K \cite{Sergey}.  Consequently we believe that
the study of Naberezhnov {\it et al.} properly characterized the
coupled modes of PMN at small $q$, whereas the study of Gehring, Park
and Shirane characterized the modes at intermediate $q$ in which the
highly unusual waterfall was discovered.

Future measurements are being planned to determine whether or not an
applied electric field can influence the shape of the waterfall.

\section{Acknowledgments}

We thank L.\ E.\ Cross, R.\ Guo, S.\ -E.\ Park, S.\ Shapiro, N.\ 
Takesue, and G.\ Yong, for stimulating discussions.  Financial support
by the U.\ S.\ Department of Energy under contract No.\ 
DE-AC02-98CH10886 is acknowledged.  Work at the Ioffe Institute was
supported by the Russian Fund for Basic Research grant 95-02-04065,
and by the National Program ``Neutron Scattering Study of Condensed
Matter.''  We also acknowledge the support of the National Institute
of Standards and Technology, U.\ S.\ Department of Commerce, in
providing the neutron facilities used in this work.

\end{document}